\documentclass[article,preprint,superscriptaddress]{revtex4}
\usepackage{graphicx}
\usepackage{amsmath,amsfonts}

\begin{document}

\title{Efficient unidirectional nanoslit couplers for surface plasmons}

\author{F. L\'opez-Tejeira}
\affiliation{Departamento de F\'{\i}sica de la Materia Condensada-ICMA,
Universidad de Zaragoza, E-50009 Zaragoza, Spain}

\author{Sergio G. Rodrigo}
\affiliation{Departamento de F\'{\i}sica de la Materia Condensada-ICMA,
Universidad de Zaragoza, E-50009 Zaragoza, Spain}

\author{L. Mart\'{\i}n-Moreno}
\email[Electronic address: ]{lmm@unizar.es}
\affiliation{Departamento de F\'{\i}sica de la Materia Condensada-ICMA,
Universidad de Zaragoza, E-50009 Zaragoza, Spain}

\author{F. J. Garc\'{\i}a-Vidal}
\affiliation{Departamento de F\'{\i}sica Te\'orica de la Materia
Condensada, Universidad Aut\'onoma de Madrid, E-28049 Madrid, Spain}

\author{E. Devaux}
\affiliation{Laboratoire de Nanostructures, ISIS, Universit\'e Louis
Pasteur,F-67000 Strasbourg, France}

\author{T. W. Ebbesen}
\affiliation{Laboratoire de Nanostructures, ISIS, Universit\'e Louis
Pasteur,F-67000 Strasbourg, France}

\author{J. R. Krenn}
\affiliation{Institute of Physics, Karl Franzens University,
Universit\"{a}tsplatz 5, A-8010 Graz, Austria}

\author{I. P. Radko}
\affiliation{Department of Physics and Nanotechnology, Aalborg
University, DK-9220 Aalborg, Denmark}

\author{S. I. Bozhevolnyi}
\affiliation{Department of Physics and Nanotechnology, Aalborg
University, DK-9220 Aalborg, Denmark}

\author{M. U. Gonzalez}
\affiliation{Laboratoire de Physique de l' Universit\'e de
Bourgogne, UMR CNRS 5027, F-21078 Dijon, France}

\author{J. C. Weeber}
\affiliation{Laboratoire de Physique de l' Universit\'e de
Bourgogne, UMR CNRS 5027, F-21078 Dijon, France}
\author{A. Dereux}
\affiliation{Laboratoire de Physique de l' Universit\'e de
Bourgogne, UMR CNRS 5027, F-21078 Dijon, France}

\maketitle

{\bf Plasmonics is based on surface plasmon polariton (SPP) modes
which can be laterally confined below the diffraction limit,
thereby enabling ultracompact optical
components\cite{barnes03,ozbay06}. In order to exploit this
potential, the fundamental bottleneck of poor light-SPP coupling
must be overcome. In established SPP sources (using
prism\cite{prism,lamp01apl}, grating\cite{grating} or
nanodefect\cite{defect} coupling) incident light is a source of
noise for the SPP, unless the illumination occurs away from the
region of interest, increasing the system size and weakening the
SPP intensity. Back-side illumination of subwavelength apertures
in optically thick metal
films\cite{Feldman,devaux03,yin04,popov05,agra05,lalanne05,Chang}
eliminates this problem but does not ensure a unique propagation
direction for the SPP. We propose a novel back-side
slit-illumination method based on drilling a periodic array of
indentations at one side of the slit. We demonstrate that the SPP
running in the array direction can be suppressed, and the one
propagating in the opposite direction enhanced, providing
localized unidirectional SPP launching.}

A picture of the proposed SPP-launcher is shown in Figure 1. A
periodic array of one-dimensional (1D) indentations is fabricated at
the (output) metal surface close and parallel to the illuminated
slit. The design of this device is based on two facts. The first one
is that the reflection of SPPs by a periodic array of indentations
presents maxima at the low-$\lambda$ edges of the plasmonic bandgaps
\cite{bozhe05oc1, flt05}. For subwavelength indentations, the
spectral locations of these edges can be obtained by folding the
dispersion relation of SPPs for a {\it flat} metal surface into the
first Brillouin zone, satisfying the following expression:

\begin{equation}
k_p P= m\pi, \label{eqRmax}
\end{equation}
where $P$ is the period of the array, $k_p$ holds for in-plane plasmon wave-vector and $m$ is the
band index. Remarkably, although the reflectance maxima depends on
groove geometry (width and depth) and number of grooves, their
spectral locations do not.

The second fact is that the phase picked up by the SPP upon
reflection is just $m \pi$, precisely at the condition given by
Eq.(1), as obtained by the modal expansion developed in Ref.
\cite{flt05}.  Using these two results, a very simple scheme for the
efficient unidirectional launching of SPPs can be envisaged. For a
given frequency, by choosing $P$ such that condition given by
equation (1) is fulfilled, a SPP emerging from the slit to the left
side will be mainly back-scattered. The interference of this
reflected SPP with the one leaving the slit to the right can be
tuned by adjusting the separation, $d$, between the slit and the
first groove of the array (defined centre to centre).
The total phase difference, $\phi$,
between the two interfering SPPs will be the phase picked up upon
reflection plus the one associated to their different path lengths
along the metal:
\begin{equation}
\phi=2k_p d +m\pi \label{eqphi}
\end{equation}
According to Eq.(\ref{eqphi}), destructive or constructive
interference should occur for those $\phi$-values equal to odd or even
multiples of $\pi$, respectively. In these latter cases, the
device would behave as an efficient source for unidirectional SPPs.

Note that Eq.(2) is based on two main simplifications.
Firstly, the previous discussion is based on the reflection of SPPs by
a groove array, while the electromagnetic (EM) fields radiated by
the slit are, at short distances, more complex\cite{flt05}.
Secondly, Eq.(2) does not take into account the radiation from the
grooves back into the slit while, in principle, EM-fields at all
openings should be self-consistently calculated\cite{LMM03}.

In order to check the validity of Eq.(2) we have carried out
numerical calculations by means of both modal expansion \cite{flt05}
and Finite-Difference-Time-Domain (FDTD) \cite{fdtd} methods. FDTD
is virtually exact for this type of 1D structures, as very small
grid sizes can be used. On the other hand, the modal expansion
treats only approximately the finite conductivity of the metal but
provides a very compact representation for the EM fields, favouring
the physical interpretation and, in some simple cases, the
calculation of analytical expressions. We characterize the
efficiency of the slit+groove system as SPP-launcher by the
"enhancement factor", $F_R$, defined as the quotient between the
current intensity of right propagating SPP ($J_R$) with and without
the grooves. Strictly speaking $F_R$ provides the efficiency of the
output side of the device; the total efficiency, defined as the
percentage of laser beam energy transferred onto the plasmon
channel, depends also on the lateral beam size, dielectric constant
of the substrate, width of the metal film, corrugation on the input
side, etcetera. Notice also that $F_R>2 $ implies that, in the
corrugated structure, the right propagating SPP carries more current
than the total SPP current (left- plus right- moving) in the single
slit case so, some or the power radiated out of plane is redirected
onto the SPP channel.

The model system is a nano-slit SPP-launcher perforated on a gold
film\cite{Vial}, designed to operate at a wavelength of $800$ nm,
inside the near infra-red range of EM spectrum. We consider an array
of $10$ grooves with a period $P=390$nm, obtained from Eq.(1) with
$m=1$. The depth of the grooves is chosen to be $w=100$nm, while the
width of both grooves and slit is $a=160$nm, which are typical
experimental parameters. Figure 2 renders the calculated (modal
expansion: black curve, FDTD: blue curve) dependence of $F_R$ with
distance $d$. In this figure, vertical lines mark the locations of
maximum interference predicted by Eq.(2). The agreement between the
modal expansion and FDTD results is excellent, except for the
behaviour at very short distances ($d \approx 2a$), due to the cross
coupling between the slit and the first groove through the vertical
walls, which is neglected within the modal expansion. More
importantly, the locations of maximum $F_R$ are accurately predicted
by Eq.(2), which allows us to design SPP-launchers without elaborate
numerical calculations.

Notice that $F_R$ would be $4$ if the whole amplitude of the
left-going SPP could be added constructively to the right-going one,
while in our simulations a smaller value is always obtained.
Calculations with the modal expansion show that this is due to the
out-of-plane scattering of the left-going SPP by the array of
grooves. The effect on $F_R$ of both damping across the flat gold
surface and partial transmission across the finite array plays a
very minor role for the considered parameters.

In order to test experimentally our proposal, several samples were
prepared with a Focused Ion Beam (FIB) in $300$nm-thick gold films
for different values for $d$, all other geometrical parameters being
the same as in the previous calculations.  Each sample consists of a
single long slit flanked by a finite periodic groove array which
extends over only half of the slit length (see Fig. 1a). This sample
design allows the quantitative experimental study of the SPP
launching efficiency, as the "isolated" slit (upper part) can be
used as an in-chip reference. The set of samples was imaged by a
Photon Scanning Tunneling Microscope (PSTM) making use of an
incident focused beam illumination for frequencies in the
[$765,800$]nm interval. Due to specific features of the experimental
set-up used for measurements in the optical regime, the incident
laser beam was directed on the sample attached to a prism under an
angle of $43^0$ with respect to the normal. Notice, however, that the
choice of angle of incidence is not critical for the spatial
distribution of transmitted energy, as a subwavelength slit in an
optically thick metal film transmits only the fundamental mode. For
each distance $d$, a pair of images was recorded by scanning at a
constant distance of about $60$ to $80$ nm from the sample surface.
The first image of the pair, corresponding to the SPP launching by a
single slit, is obtained by focusing the laser beam on the upper
part of the slit. For the second image, the laser beam is moved to
the lower part in order to collect the data for the slit+grating
case. Image pairs for $d=585$nm and to $d=486 $nm are displayed in
Figure 2b (left panels: single slit, right panels: slit+grating).
Figure 2b clearly shows that the grating enhances the intensity of
the right propagating SPP for $d=585$nm whereas for $d=486$nm this
intensity is greatly reduced. In order to quantify this enhancement,
an average longitudinal cross-cut of each image is obtained by using
$20$ longitudinal cross-cuts, corresponding to different coordinates
along the slit axis. Then, the relative position of the two average
cross-cuts composing each image pair is adjusted so that the
saturated areas (i.e. the signal taken right on top of the slit) are
super-imposed. Finally, the experimental enhancement factor, $F_R$,
is extracted by averaging the ratio between the two curves along the
longitudinal cross-cut. Figure 2a renders experimental results
(squares) for $F_R$ for the five different samples fabricated.
Experimental data are in good agreement with the theoretical
predictions, definitely showing that the presence of the grating
modulates the coupling into the right propagating SPP. We find this
agreement quite remarkable, especially when taking into account that
each experimental point in Fig. 2a corresponds to a different
sample.

We have also designed similar samples for efficient unidirectional
SPP excitation at telecom (TC) wavelengths, increasing
correspondingly the grating period and its separation from the slit.
In this case, normal incidence back-side illumination is allowed by
the experimental set-up and used in all experiments. The SPP
propagation length is significantly longer in this wavelength range
($\approx 200 \mu m$) rendering a simple way of determining the
enhancement factor. Similarly to the experiments described above,
this factor has been obtained by using two near-field optical images
taken at two positions of the focused laser beam incident normal to
the sample surface (from the back side). The enhancement was simply
determined as a power ratio between the SPP beams excited for two
different adjustments, the SPP beam power being estimated far away
($\approx 50 \mu m $) from the slit. The typical near-field optical
image obtained when illuminating the lower slit part (adjacent to
the grating) is shown in Fig. 3 (upper panel), featuring a strong
SPP beam propagating away from the slit in the direction opposite to
that of grating and demonstrating thereby the desirable effect of
the enhanced unidirectional SPP excitation.

It should be mentioned that the enhancement factor determined in
this way exhibited a significant dispersion due to inaccuracy in the
illuminating laser beam adjustment.
Consequently, several series of measurements were performed
conducting independent adjustments for each sample and wavelength.
Averaged results and estimated errors are shown in Fig. \ref{fig3},
which shows the wavelength dependence of the SPP excitation
enhancement caused by the grating array. As can be seen, the
comparison between theory and experiments is rather satisfactory:
for the case of the sample with $d=P+P/2=1125$ nm, the enhancement
factor decreases as the wavelength increases (with the only
exception of a sharp peak at 1520 nm), evolving from an enhanced
regime ( $F_R \approx 2$) to one in which SPP coupling is clearly
diminished by the grating ( $F_R <1$). On the other hand, $F_R
\approx 2$ all over the range for the sample with $d=P-P/4=562$ nm,
as predicted by the modal expansion calculation.

Another application of the system proposed in this paper is the
focusing of SPPs, creating local field enhancement ("hot spot") at
a given location. Focusing of SPP has been achieved through the
interaction of SPP with curved surface
corrugations\cite{Nomura05,Yin05,Liu05,Offerhous05,steele06}. The
proposed approach for the localized unidirectional excitation of a
SPP beam can be generalized to focus SPPs with a higher efficiency
than in the cited study while, at the same time, blocking the
propagation away from the focus. Importantly, in these curved
structures, the rigorous modelling of the SPP excitation needed
for optimization of the focusing would be difficult, while the
relation given by Eq.(2) still provides simple design rules. As a
proof of principle, in this paper we present focusing at telecom
wavelengths by milling a curved slit along with the corresponding
grating grooves [Fig. 4(a)], using a similar configuration to that
illustrated with the near-field optical image shown in Fig. 3. The
effect of SPP beam focusing was clearly seen already at the stage
of far-field adjustment (using a microscope arrangement with an
infrared CCD camera) due to weak out-of-plane SPP scattering by
surface roughness [Fig. 4(b)]. The typical near-field optical
image obtained at the wavelength of 1520 nm demonstrates efficient
focusing of a launched SPP beam at the center of slit curvature
with a spot size of $~ 2.5 \times 2.5 \mu m^2$ [Fig. 4(c)]. More
generally, the design of other more complicated curved structures
based on these principles can be envisaged allowing, for example,
the excitation of SPP beams propagating in different directions
and focused at different locations. Further investigations in this
direction are being conducted.

In conclusion, we have studied the SPP launching by subwavelength
slit apertures with back-side illumination, demonstrating a novel
method allowing for efficient localized unidirectional generation by
means of a finite grating of grooves adjacent to the slit. Our
simple model enables us to make quantitative predictions that have
been experimentally confirmed for both near infrared and telecom
ranges. Such analytical predictions on coupling-enhancement have
also been found in good agreement with sophisticated computer
simulations, irrespective of our model's simplified description of
some of the physics involved. With respect to practical
applications, we have shown that the SPP coupling-in can be enhanced
by a factor of $3$ without any modification on the illumination
source. This is particularly relevant if we keep in mind that the
total transmitted current can also be easily enhanced making use of
a symmetric grating at the illuminated surface \cite{fj03prlmay}.
Moreover, the SPP beam is confined only at one side of the aperture
and it can also be suppressed with a careful selection of the
geometrical parameters. We have also shown how a slight modification
of the system can be used to concentrate freely propagating SPP's in
region with lateral dimensions comparable to the SPP wavelength. In
our opinion, the possibility of efficient local SPP excitation
producing a well defined and collimated SPP beam is of great
importance and extremely useful for various plasmonic devices of
current interest.

\begin{figure}[p]
\includegraphics[width=15cm]{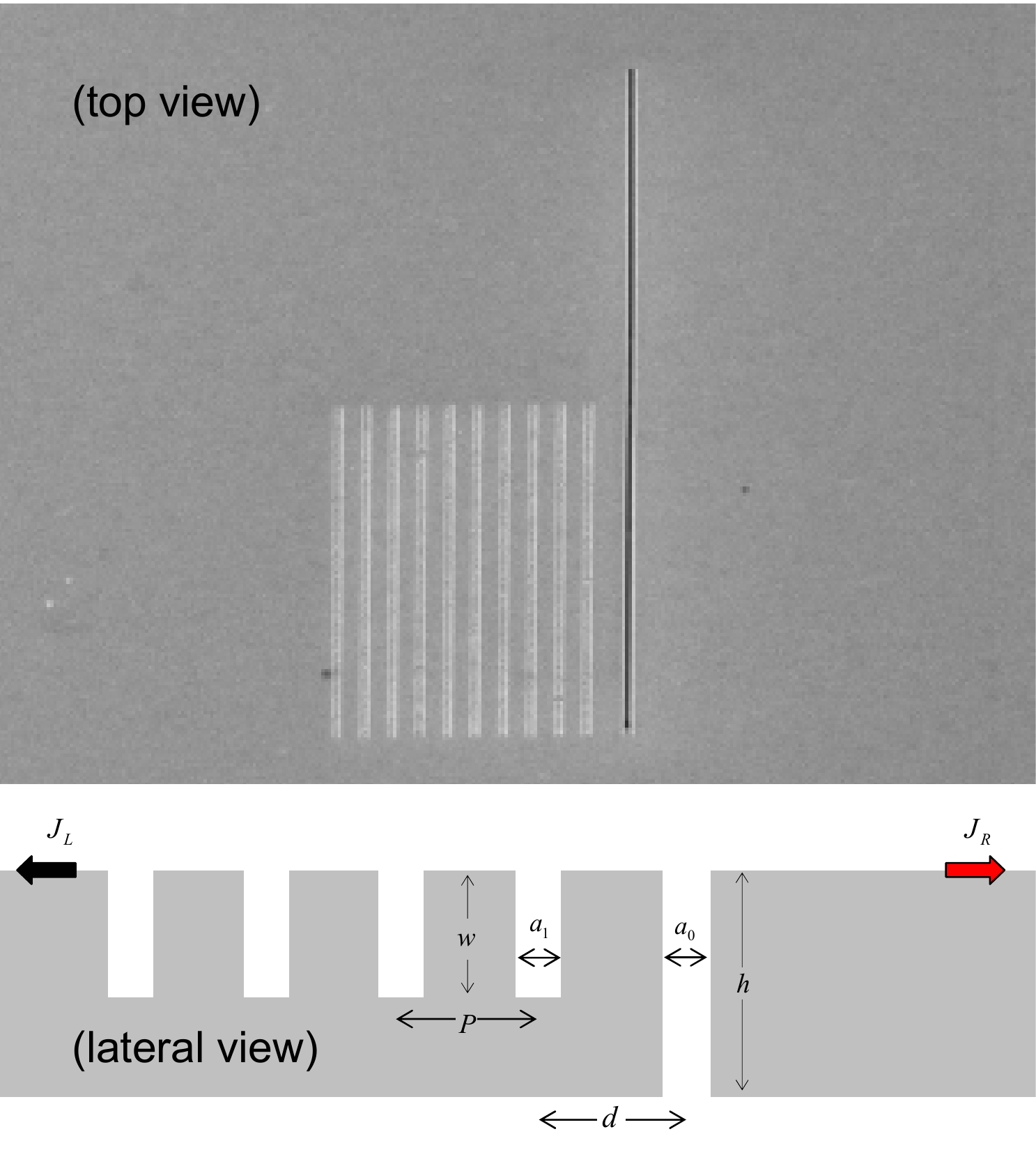}
\caption{SEM and schematic pictures of the structures investigated.
Parameters used in the definition of the slit, grooves and metal
film are also shown. $J_R$ and $J_L$ stand for the current energy
densities for right- and left- propagating surface plasmons,
respectively}\label{fig1}
\end{figure}

\begin{figure}
\includegraphics[width=15cm]{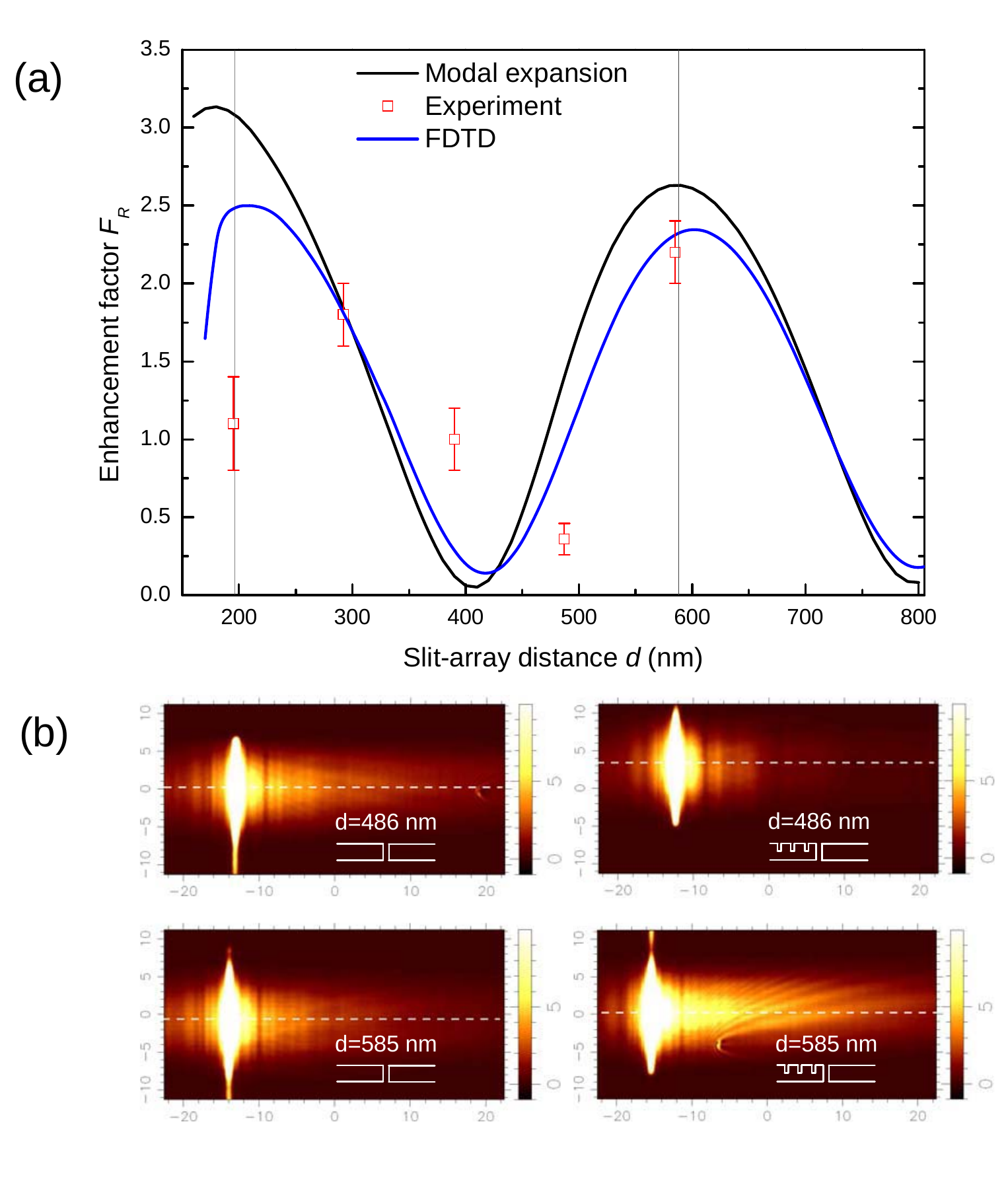}
\caption{\textbf{(a)} Dependence with slit-array distance of the
enhancement factor $F_R$. The working wavelength is $\lambda=800$nm,
while the geometrical parameters defining the system are: slit
length is $30\mu m$, slit and groove widths $a=160$nm, groove depth
$w=100$nm and array period $P=390$nm. The figure renders the
computed results obtained by both modal expansion (black line) and
FDTD methods (blue line). Vertical lines mark the positions of the
enhancement maxima according to Eq. (\ref{eqphi}). Experimental
results are represented by squares. \textbf{(b)} PSTM images
recorded at $\lambda=800$nm for two different slit-grating
distances. Left column: isolated slit. Right column:
slit+grating.}\label{fig2}
\end{figure}

\begin{figure}
\includegraphics[width=15cm]{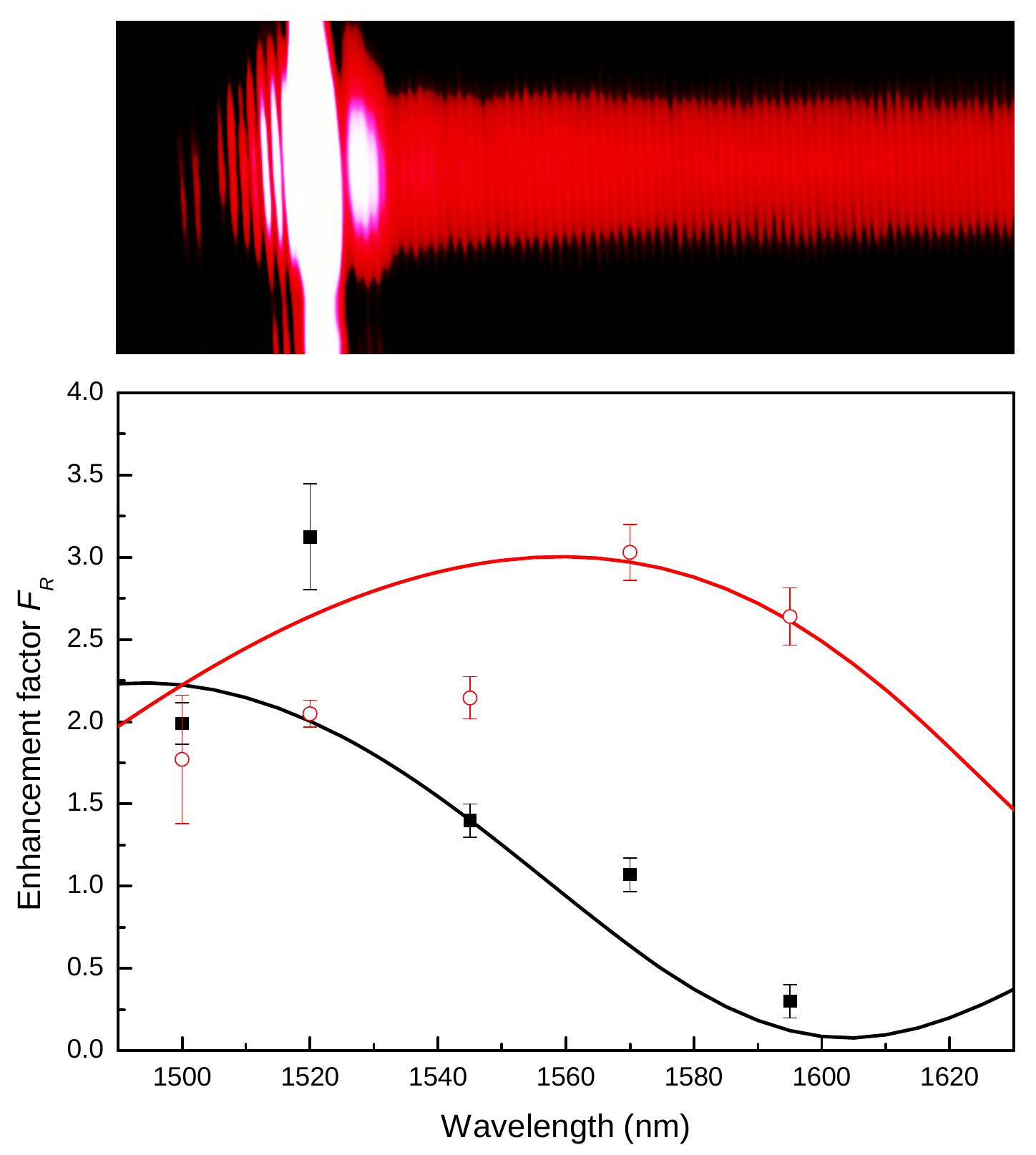}
\caption{Experimental results and modal expansion calculations for
the spectral dependence of the enhancement factor at the telecom
range. Two samples were considered. In both, the slit length is $50
\mu m$, slit and grooves widths are $400$ and $200$nm, respectively,
with groove periodicity $P=750$nm, film thickness $h=300$nm and
groove depth $w=100$nm. In one sample the slit-grating distance
$d=3P/2=1125$nm (experiment: black squares, theory: black curve)
while, in the other $d=3P/4=562$nm (experiment: red circles, theory:
red curve) The image in the upper panel depicts grating induced
enhancement for $d=3P/2$ at $\lambda=1520$nm. (Size $=70x26 \mu
m^2$). Weak periodic modulation of the strong SPP beam propagating
to the right (upper panel) is due to its interference with a SPP
wave reflected by a remote auxiliary structure.}\label{fig3}
\end{figure}

\begin{figure}
\includegraphics[width=15cm]{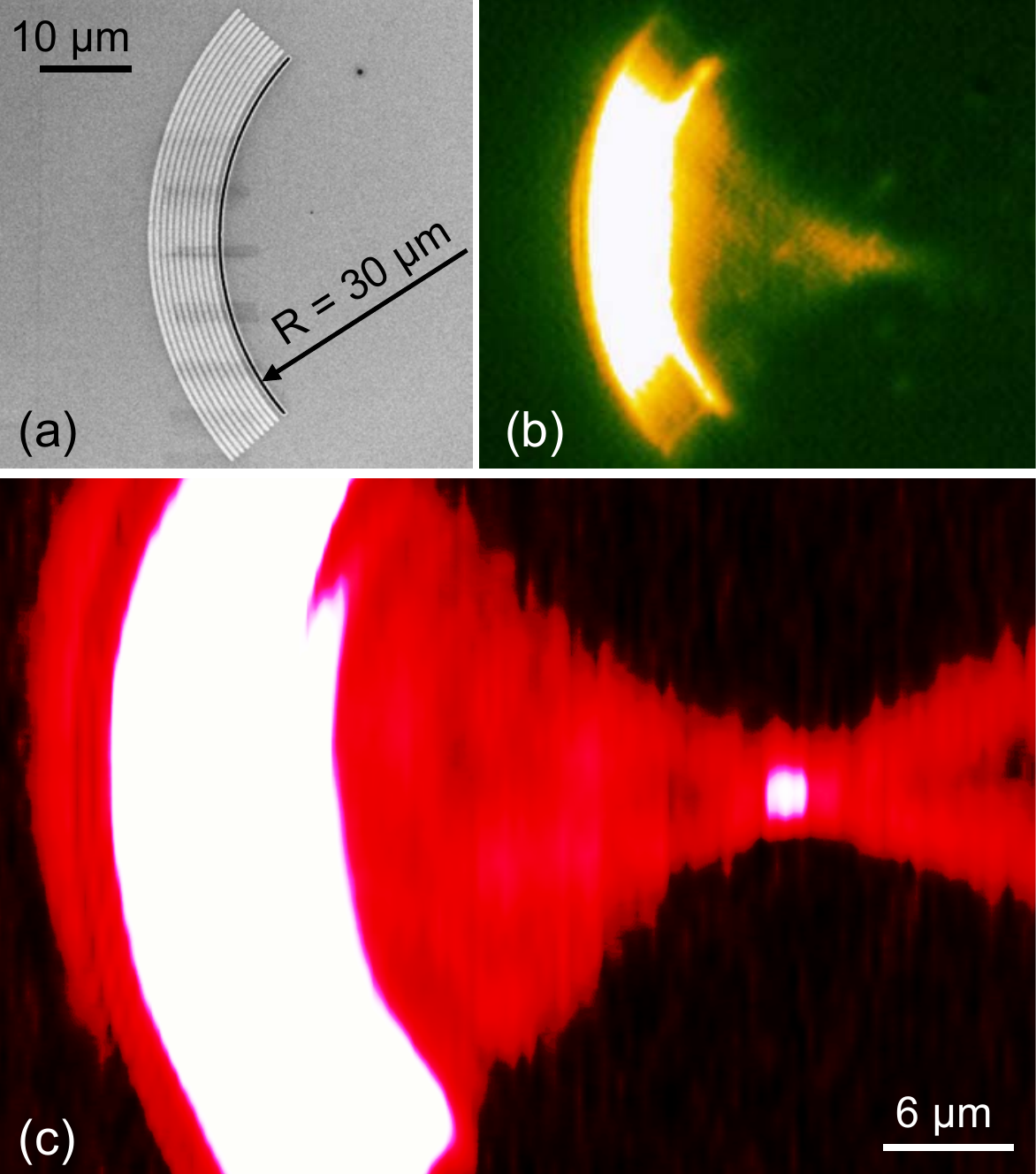}
\caption{ Demonstration of the simultaneous SPP excitation and
focusing using a curved slit flanked with concentric periodic
grooves. (a) SEM image of the fabricated structure characterized by
the slit and groove widths of 400 and 200 nm, respectively, groove
periodicity of 750 nm, groove depth of 100 nm and slit-groove
distance of 1125 nm. Film thickness is 280 nm and curvature radius
is 30 $\mu$ m. The slit chord length is $40 \mu m$. (b) Far-field
image recorded with a CCD-camera and (c) the corresponding
near-field optical image (size = $48\times 32 \mu m^2$), both being
obtained when the curved slit was illuminated at normal incidence
with radiation at the telecom wavelength of 1520 nm.}
\end{figure}

{}

\section{Acknowledgements}
Financial support by the EC under Project FP6-2002-IST-1-507879
(Plasmo-Nano-Devices) is gratefully acknowledged. We thank J.
Dintinger and J.-Y. Laluet for technical assistance.

\section{Competing financial interests}

The authors declare that they have no competing financial interest.

\end{document}